# Unconventional Localization Prior to Wrinkles and Controllable Surface Patterns of Film/Substrate Bilayers Through Patterned Defects in Substrate

Xiangbiao Liao[1], Liangliang Zhu[1], Hang Xiao[1], Junan Pan[1], Feng Hao[1], Xiaoyang Shi[1], Xi Chen[1,2]

[1]Department of Earth and Environmental Engineering, Columbia University, New York, NY 10027, USA

[2]School of Chemical Engineering, Northwest University, Xi'an 710069, China

**Abstract**

A novel bilayer is introduced, consisting of a stiff film adhered to a soft substrate with patterned holes beneath the film/substrate interface. To uncover the transition of surface patterns, two-dimensional plane strain simulations are performed on the defected bilayer subjected to uniaxial compression. Although the substrate is considered as the linear elastic material, the presence of defects can directly trigger the formation of locally ridged and then folding configurations from flat surface with a relatively small compressive strain. It is followed by the co-existing phases of folds and wrinkles under further overall compression. This phase transition reverses the traditional transition of wrinkle-to-ridge/fold for defect-free substrates. It's also found that the onset of initial bifurcation is highly dependent on the spatial configuration and geometries of holes, since the interaction of defects allows more strain-relief mechanisms beyond wrinkling. Furthermore, a rich diversity of periodic surface topologies, including overall waves, localizations, saw-like and co-existing features of folds and wrinkles can be obtained by varying the diameter, depth and spacing of holes as well as compressive strain, which provides a potential approach to engineer various surface patterns for applications.

**Introduction**

A rich range of surface morphologies, from overall wrinkles to local bumps or channels, can be often observed in natural and biological systems.[1-3] The controllability of such features allows not only extensive applications in optics[4] and flexible electronics[5], but also tuning surface adhesion[6] and hydrophobicity,[7] etc. The sinusoidal pattern of wrinkles forms in the stiff film/soft substrate bilayer subjected to uniaxial compression.[1] Anisotropy of membrane force can result in more intriguing wrinkling patterns including stripes, herringbones and zigzag labyrinths on planar substrate,[8,9] and buckyball-like and labyrinth features on spheroidal substrate.[10-12] While overall wrinkles are sufficient for some applications, dynamically and locally tunable surface features are also crucial.

Further compression induces the transition from the sinusoidal to periodic-doubling patterns, ultimately followed by localized folds in the bilayer.[13] It has been also found that pre-compression applied to the substrate can facilitate the formation of folds, while a different advanced mode of instability, termed localized mountain ridge, occurs after wrinkling if the substrate is subjected to pre-stretch.[14,15] In addition, the transitions from wrinkles to deep folds and large aspect ratio of ridges have been experimentally observed.[16,17] However, the highly non-linear elasticity of substrate determining the difference of advanced localizations limits the choice of substrate materials, and pre-compression/stretch experimentally also complicates the fabrication. Furthermore, both of ridge and fold modes appear after the formation of wrinkles at relatively high compressive strain.[16,18] Although locally controllable deformations were achieved in the particle-enhanced soft material without film attachment,[19] it is urgent to develop more flexible techniques to directly create reversible local surface features in film/substrate bilayers.

Despite these advances, rare efforts of modeling take defects in materials into account. Defects are experimentally inevitable and likely to affect the formation of wrinkles or localizations.[20, 21] Among the limited literatures, intensive studies focused on elastically heterogeneity of thin film including patterned holes[22, 23] and rigid elements[24, 25] where the wrinkling pattern is disordered in the distance scaling with characteristic wavelength. However, the influence of defects inside the substrate on controlling the surface patterns of bilayer has not received much attention. Here, we focus on the defected substrate with pre-patterned holes below the interface of film/substrate. Finite element calculations are conducted to investigate how the spatial and geometrical variation of defects impacts the evolution of surface patterns when the bilayer is subjected to uniaxial compression. The presence of defects as strain relief can induce an unconventional transition from localized ridges to folds, followed by wrinkles. We further study the dependence of initial bifurcation and advanced patterns on the defect size, depth as well as spacing.

**Method**

The representative segment of defected bilayer is shown in Figure 1, consisting of thin film adhered to a substrate with patterned holes. Periodic boundary conditions in $x$ direction are applied to the vertical sides of the model. Neither vertical displacement nor shear traction is allowed on the bottom surface. Here, nonlinear numerical simulations are performed through commercial software ABAQUS and four-node plane strain quadrilateral elements are used. In all cases, the mesh density is verified by mesh convergence studies. The film and substrate with linear isotropic elasticity have different Young's modulus, $E_f/E_s = 300$, while the Poisson's ratio is taken to be the same for simplicity. Although we focus on 2D morphological designs, our design concept can be easily extended to 3D system. In these calculations, a linear perturbation analysis is first conducted to identify the onset of first bifurcation including the morphological mode. To promote the initial instability mode, this morphology with a small amplitude, 0.01 times the film thickness, is taken as the initial geometric imperfection. We adopt pseudo-dynamic method to study advanced instability transitions within the frame of quasi-static deformation and kinetic effects are negligible. Additionally, the bilayer is subjected to the compressive strain $\varepsilon_a$, which is achieved by setting higher expansion rate in the film since the constraint from substrate makes the film in compression.

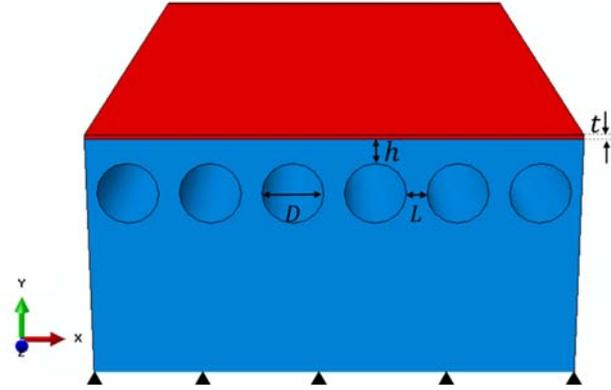

**Figure 1**. A representative segment of thin film (red) adhered to defected substrate (blue) with pre-patterned holes. $t$ is the thickness of film. $D$ and $h$ denote the diameter and depth of defects, respectively. $L$ is the nearest spacing between two adjacent defects.

**Results and Discussion**

For defect-free substrate, the critical compressive strain $\varepsilon_{wc}$ and wrinkling wavelength $\lambda_{wc}$ are calculated by[1]

$$\varepsilon_{wc} = \frac{1}{4}\left(3\overline{E_s}/\overline{E_f}\right)^{2/3} = 1.16\% \tag{1}$$

$$\lambda_{wc} = 2\pi t\left(\overline{E_f}/\overline{E_s}\right)^{1/3} = 29.0t \tag{2}$$

where the bilayer has the ratio of Young's modulus $E_f/E_s = 300$. In the defected bilayer shown in Figure 1, the spacing $L$ between defects with diameter $D$ and depth $h$ is described as the closest distance between two adjacent holes, and we define the initial bifurcation with the critical strain $\varepsilon_c$, where flat surface transits to uneven pattern, but not limited to wrinkles. To avoid boundary effect, we set the thickness of substrate as $10\lambda_{wc}$.

To study the effect of defects on the evolution of surface patterns, the defected layer with $D = \lambda_{wc}$, $h = 0.1\lambda_{wc}$ and $L = 5\lambda_{wc}$, taken as an example, is subjected to uniaxial compression. Shown in Figure 2, the surface pattern transits from flat surface to local bumps, mountain ridges, folds and the co-existing phase of folds and wrinkles with increasing compressive strain. Once the initial bifurcation occurs at the critical strain 0.62%, the flat surface locally buckles up at the location right above defects. After that, the film starts to pull material outward from the substrate, resulting in the formation of ridges. However, it's hard to achieve mountain ridges without the conditions of non-linear elasticity for substrate and pre-stretch.[15] Additionally, the critical value lower than that predicted by Equation (1) for the formation of wrinkle at the defect-free bilayer, and it means the bilayer with defected substrate is easier to relax the stress by local buckles rather than overall wrinkles. At the compressive strain of 6.0%, the mountain ridges snap downward and evolve into the folding pattern since ridges can no longer energetically favorable. Finally, another strain-relief mode, wrinkles, emerges in the flat region between two adjacent hole, leading to the co-existing phase of both folds and wrinkles, which have the characteristic wavelength determined by Equation (2). Interestingly, the wrinkles appearing far after the localizations, ridges and folds in turn, unconventionally reverse the traditional wrinkle to ridge/fold transition due to the presence of defects in the substrate.[13]

Although unavoidable small holes are imperfections of in most systems,[20, 21] the pre-patterned defects in the substrate of bilayers can be well utilized. First, the presence of holes can easily trigger the localizations at relatively small compression 0.62%, while the ridges/folds occurs usually with much large loading as well as the condition of pre-compression/stretch. Furthermore, the strain for wrinkles occurring, approximate 18%, is much larger than that in defect-free system (1.16%), which underpins a potential method by holes to suppress wrinkles regarded as a failure mechanism in engineering systems.

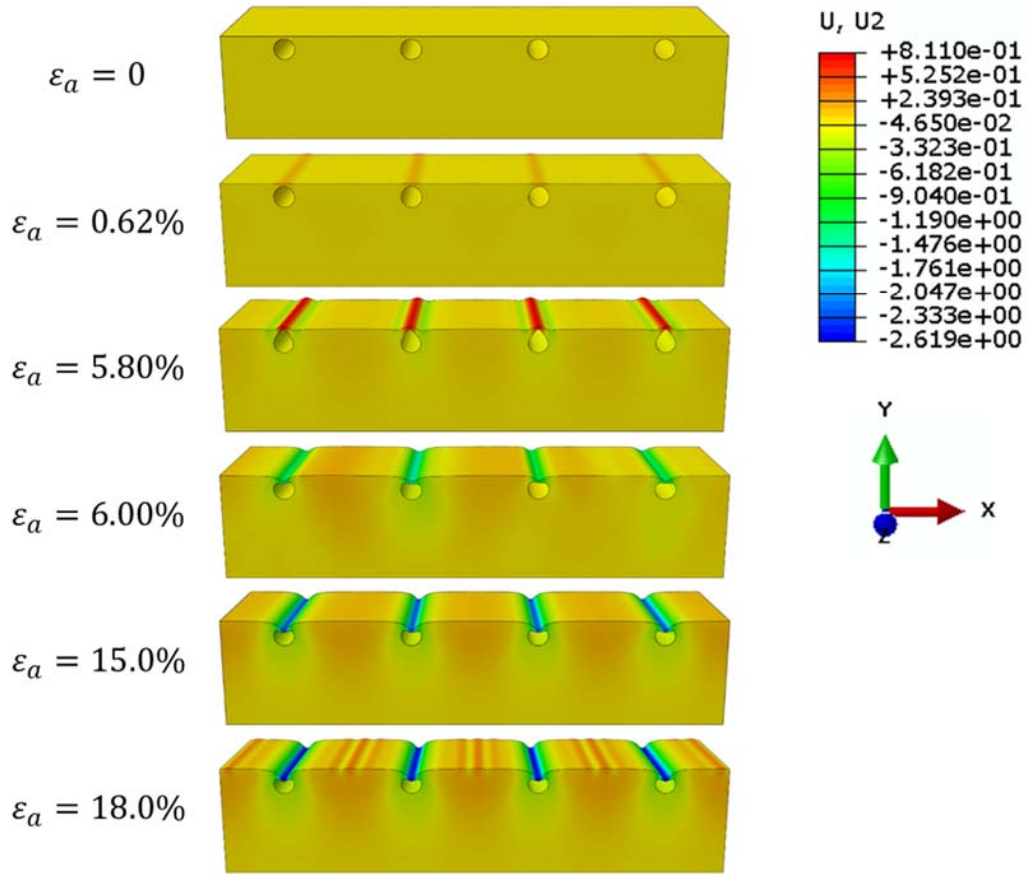

**Figure 2**. Morphological evolution of the defected bilayer structure subjected to increasing compressive strain, where $D = \lambda_{wc}$, $h = 0.1\lambda_{wc}$ and $L = 5\lambda_{wc}$

In order to study how the spatial configuration of defects on morphological evolution, we calculate the critical strain for initial mode of instability as a function of normalized spacing, illustrated in Figure 3. It's found that defected bilayer bifurcate at smaller compressive strain than that for defect-free one of 1.21%, which is approximately consistent with the theoretical results in Equation (1). Shown by the serial cases of $D = \lambda_{wc}$ and $h = 0.1\lambda_{wc}$ (black curve), the critical strain increases as the spacing $L$ becomes larger up to a platform. The reason for increasing critical strain lies in less interaction between two adjacent defects as the spacing increases, thus the mechanism of initial bifurcation varies with the spacing $L$. It's evident that the pattern of waves occurs if the spacing is small, while large spacing only allows locally buckling up. From the bottom insets of Figure 2, the transformation of one dimensional hole lattice into mutually orthogonal and alternating pattern induces the surface pattern having spacing doubling wavelength.

The position and size of defects can also affect the initial buckling mode. Take the examples with fixed diameter $D = \lambda_{wc}$, it's observed that the critical strain becomes larger when defects locate deeper relative to the film/substrate interface. When the depth is beyond half of characteristic wavelength $\lambda_{wc}$, the critical strain approaches the value for wrinkling in non-defect system since negligible effect of defects on surface pattern. In addition, larger defects result in smaller critical strain, although the critical strains are also increasing functions with regard to spacing. If the size is very small, there is no pronounced influence on wrinkling pattern. Through the parametric studies, the initial instability for defected bilayer shows big difference with the perfect bilayer unless the defect size is much smaller than $\lambda_{wc}$ or the defects locate far from the film/substrate interface.

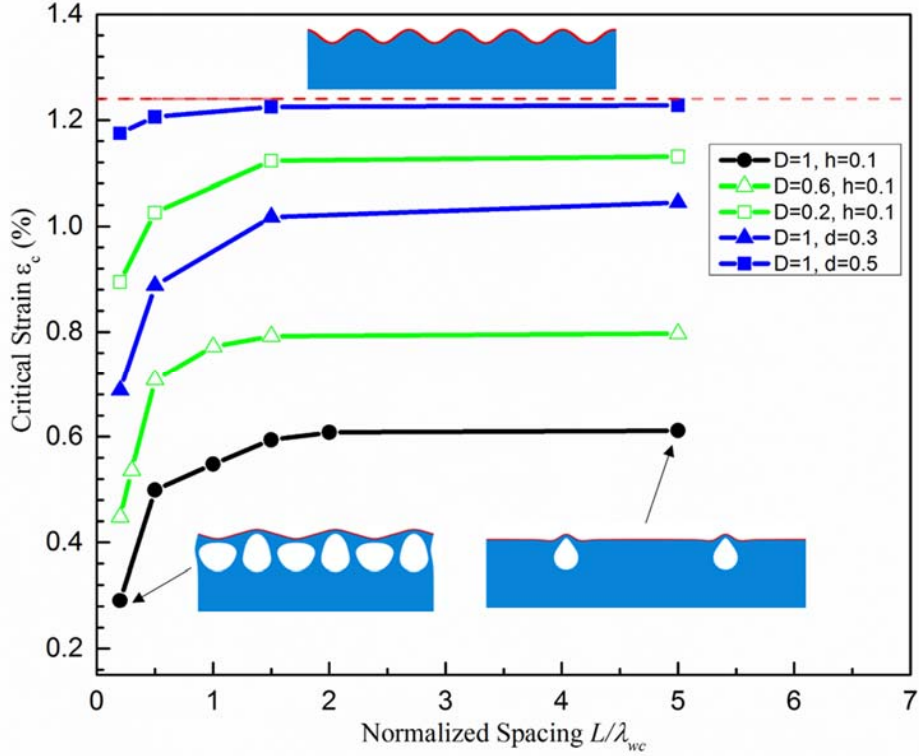

**Figure 3**. The strain for first bifurcation as a function of spacing $L$ in the bilayer with defects having different diameters and depths. The top inset is wrinkling pattern for non-defect structure, and the bottom insets are buckling morphologies for the defected bilayers with the same $D = \lambda_{wc}$, $h = 0.1\lambda_{wc}$, but different normalized spacing $L/\lambda_{wc}$.

After initial mode of instability, the global compression can be further increased. Since the spacing between holes plays a significant role in bifurcation of defected layers, we set the size of depth of holes a constant. The diagram of advanced surface pattern is plotted with regard to applied strain $\varepsilon_a$ and normalized spacing $L/\lambda_{wc}$ in Figure 4. When two adjacent holes are relatively close, the amplitude of waves builds up with increasing applied strain after the initial bifurcation. The continuously applied compression induces locally ridged pattern followed by the folding and wrinkling configuration for the case of large spacing, for example, $L = 5\lambda_{wc}$. However, if the spacing is $L = \lambda_{wc}$, we observe the saw-like pattern, which possibly results from the interaction between locally buckling up and overall wave. With further compression, the saw-like pattern evolves into the pattern with partial folds.

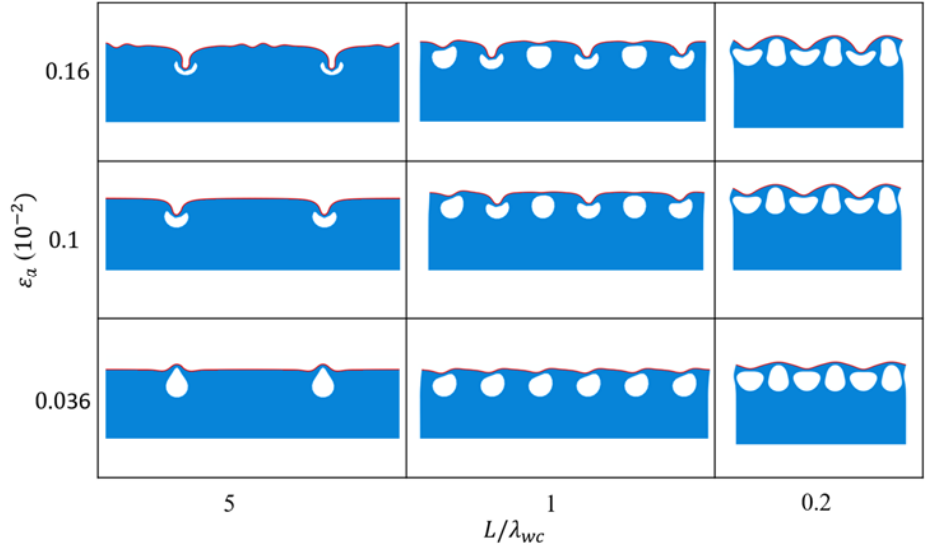

**Figure 4**. The diagram of buckling morphologies for the defected structures ($D = \lambda_{wc}$ and $h = 0.1\lambda_{wc}$) against applied strain and normalized spacing.

Figure 5 shows the diagram of surface patterns against normalized diameter $D/h$ and normalized spacing $L/\lambda_{wc}$. While $h$ may also be variable, all results presented in this diagram are pretty similar as long as $h$ is fixed. All the morphologies are captured at the strain of 2.0%, slightly larger than $\varepsilon_{wc}$. When the diameter is relatively small, we observe the roughly wrinkling patterns little perturbed by the defects inside substrate. Note that perfect wrinkling morphology are expected in non-defects systems ($D/h \to 0$). However, in the case of $D/h = 2$ and $L/\lambda_{wc} = 0.2$, the wavelength of waved pattern is spacing-tripling and it's a new instability configuration due to defects interaction. Based on these diagrams, the defected bilayer system can bifurcate into diverse modes: localized ridges, localized folds, saw-like and waved patterns, dependent on the size, position and spacing of defects.

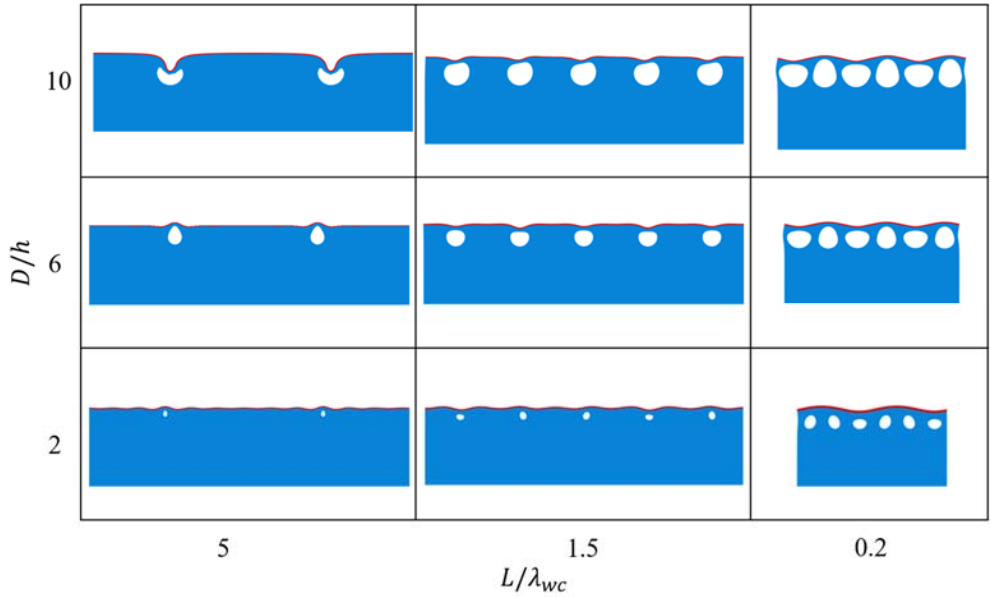

**Figure 5**. The diagram of surface morphologies for the defected structures ($h = 0.1\lambda_c$), where the applied strain is fixed at $\varepsilon_a = 0.02$, against normalized diameter $D/h$ and normalized spacing $L/\lambda_{wc}$

Despite of 2D simulations in this study, we can easily extend the work to 3D systems, where large freedoms of defects geometries, anisotropy of loading and instability mechanisms possibly further enrich the family of surface pattern in the bilayer. Not limited to the method of mismatch expansion to apply compressive strain in the film, there are other practical mechanisms including mechanical compression, active material responding to environmental stimulus, *e.g.*, pH, electrical field and humidity. The effect of non-linear elasticity of materials, for example, incompressible neo-Hookean materials, will be subjected to the future work.

## Conclusion

In this study, we show that the conventional transition of wrinkles to localizations (folds or ridges) in film/substrate bilayers, subjected to increasing uniaxial compression, can be reversed through the pre-patterned defects in the substrate. Compared with defect-free bilayers, the localizable surface topologies are achieved at relatively small strain even with linear-elastic substrate. The presence of holes not only simplifies the conditions for creating localized morphologies, but also serves as a strain-relief mechanism to suppress the formation of overall wrinkles, regarded as failure in some cases.

Additionally, the initial mode of instability and advanced surface patterns of defected bilayers are highly dependent on the spatial and geometric configurations of defects. The flexibility, provided by hole interactions, underpins a novel design of tunable surface patterns, including localized ridges and folds, waves with different wavelengths, saw-like features and co-existing phase of wrinkles and folds. The controlling surface patterns with defected bilayers across a wide range of scales has potential to optimize performance of diverse applications such as surface adhesion, hydrophobicity and friction.[6, 26]

## Acknowledgements

X.C. acknowledges the support from the National Natural Science Foundation of China (11172231 and 11372241), ARPA-E (DE-AR0000396) and AFOSR (FA9550-12-1-0159); X.L., L.Z. and H.X. acknowledge the China Scholarship Council for the financial support.